\documentclass[runningheads]{llncs}
\usepackage[T1]{fontenc}
\usepackage{graphicx}

\usepackage[hyphens]{url}
\usepackage[hidelinks]{hyperref}

\usepackage{color}

\begin{document}
\title{Tracing and Metrics Design Patterns for Monitoring Cloud-native Applications}
\titlerunning{Tracing and Metrics Design Patterns}

\author{Carlos Albuquerque\inst{1} \and
Filipe F. Correia\inst{1}} 
\authorrunning{C. Albuquerque et al.}

\institute{INESC TEC, Faculty of Engineering, University of Porto, Portugal\\
\email{\{up201706735,filipe.correia\}@fe.up.pt}}

\maketitle
\begin{abstract}
Observability helps ensure the reliability and maintainability of cloud-native applications. As software architectures become increasingly distributed and subject to change, it becomes a greater challenge to diagnose system issues effectively, often having to deal with fragmented observability and more difficult root cause analysis. This paper builds upon our previous work and introduces three design patterns that address key challenges in monitoring cloud-native applications.

\textsc{Distributed Tracing} improves visibility into request flows across services, aiding in latency analysis and root cause detection, \textsc{Application Metrics} provides a structured approach to instrumenting applications with meaningful performance indicators, enabling real-time monitoring and anomaly detection, and \textsc{Infrastructure Metrics} focuses on monitoring the environment in which the system is operated, helping teams assess resource utilization, scalability, and operational health.

These patterns are derived from industry practices and observability frameworks and aim to offer guidance for software practitioners. 


\end{abstract}

\section{Introduction}
\label{sec:intro}

The increasing complexity of cloud-native applications calls for advanced observability mechanisms to ensure reliability, maintainability, and performance optimization. Cloud-native software systems, often built on microservices architectures and container orchestration platforms, introduce new challenges in fault diagnosis, performance analysis, and real-time monitoring. Unlike traditional monolithic systems, these environments are highly dynamic, with ephemeral workloads, distributed state, and a continuous deployment lifecycle.

Despite the wide availability of monitoring tools and frameworks, many organizations struggle to achieve effective \textit{observability}, as traditional monitoring approaches often fall short in providing the depth of insight needed to diagnose complex, distributed systems and ensure their reliability in cloud environments. Observability extends beyond conventional monitoring by enabling developers and operators to understand \textit{why} a system behaves a certain way, rather than just identifying \textit{what} went wrong. Observability is typically achieved through three types of telemetry---\textit{logs} are immutable, timestamped records of discrete events, which are essential for understanding discrete failures and debugging unexpected behaviors; \textit{metrics} are numerical data points collected over time, often aggregated, giving insight into system health and performance trends, and enabling proactive responses to degradation and scaling issues; and \textit{traces} capture the journey of a request across distributed components, helping teams detect latency bottlenecks and pinpoint failures across service boundaries. Together, these telemetry data types empower teams to diagnose issues more quickly, track regressions, and continuously improve system stability. Telemetry data combined with the need for automated anomaly detection and root cause analysis, calls for systematic and scalable observability practices. Existing tools can come a long way in helping professionals address use such practices, but developers still need guidance on which practices are most appropriate given the context at hand and how to go about adopting them effectivelly. 

Patterns capture proven practices in a reusable format, offering structured solutions that can help engineers apply and adapt them in the real-world. Building upon our previous works~\cite{Albuquerque-2022,Albuquerque-2023,Albuquerque-2024}, this paper refines and extends observability patterns tailored for cloud-native applications. We introduce three additional patterns that address gaps in distributed tracing, application metrics collection, and infrastructure monitoring. These patterns provide practical guidance for engineers looking to enhance the observability of their cloud-native systems.

In the remainder of the paper, Section~\ref{sec:related-work} quickly reviews the related work, Section~\ref{sec:about-patterns} explains the pattern mining process and outlines the pattern catalog and its evolution, and Sections~\ref{pat:distributed-tracing}, \ref{pat:application-metrics}, and \ref{pat:infrastructure-metrics} describe the proposed patterns in detail. Finally, Section~\ref{sec:conclusions} discusses conclusions and future research directions.

\section{Related Work}\label{sec:related-work}

Many resources, from books to blog posts to academic articles, discuss best practices for developing cloud-native systems, and quite a few have been documented as patterns~\cite{Sousa-2020,Sousa2018c,Sousa2015,sousa2016engineering,Maia2022,maia2024patterns,maia2025container,Dobaj2019,cdp_aac}. Some of such resources address monitoring and observability concerns~\cite{Newman-2021,Waseem-Practices-2021,Li-2021,Darrington-2023,faseeha2025observability}, but only a fraction of these have been written in the form of patterns~\cite{Brown-2021,Richardson-Wiki-2021,Sousa2017,Sousa2018b}.

We will not go into the details of each of these works in this section, but find it worth highlighting those that are closest to our own work. In particular, Sousa et al. have described \textsc{External Monitoring}, \textsc{Preemptive Logging}, and \textsc{Log Aggregation}~\cite{Sousa2017,Sousa2018b}, Brown et al. also catalogued \textsc{Log Aggregation}, \textsc{Correlation ID} and \textsc{Query Engine}~\cite{Brown-2021}, and Richardson’s microservices pattern language contains very early drafts of some monitoring-related patterns such as \textsc{Application Metrics} and \textsc{Distributed Tracing}~\cite{Richardson-Wiki-2021}.

\section{About the Patterns}
\label{sec:about-patterns}

We started the process of mining the patterns in this paper by reviewing research literature that specifically mentioned monitoring practices. We then looked into grey literature to deepen our understanding of the practices, which proved particularly useful to find real-world accounts of the patterns' use. Lastly, we looked into tools---we first identified commonly used features of monitoring tools and then searched for existing grey literature on how those features should be used. This allowed us to further refine some of the patterns. 

\begin{figure*}[b]
\centering
\includegraphics[width=1.03\textwidth]{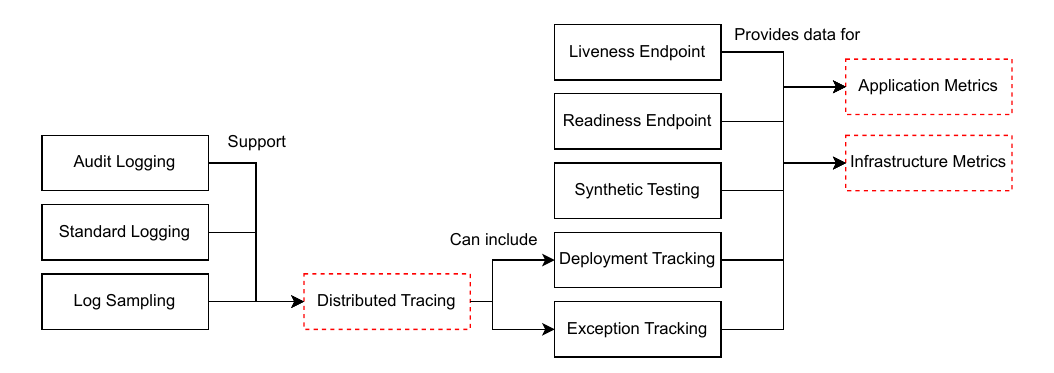}
\caption{Overview of the monitoring design pattern candidates proposed by the authors. The patterns highlighted with a dashed line are the ones explored in this paper.}
\label{fig:pattern-map}
\end{figure*}

Our previous papers~\cite{Albuquerque-2022,Albuquerque-2023,Albuquerque-2024} have already contributed different patterns to support observability in cloud-native environments. This paper builds upon these works and describes three more patterns, completing our catalog of eleven patterns for monitoring cloud-native applications~\cite{Albuquerque-Thesis-2022}. Figure~\ref{fig:pattern-map} shows a map of these patterns, and we often reference the other design patterns in the catalogue throughout the paper. The list below briefly describes each pattern, through its name, solution summary, and a reference to where the reader can find more about the pattern.

{\small
\begin{enumerate}
    \item \textsc{Audit Logging} --- Log in a data store who did what, where, and when. This data store can later be queried to reproduce issues reported by the users~\cite{Albuquerque-2024}.
    \item \textsc{Standard Logging} --- Implement logging using a consistent format across all services~\cite{Albuquerque-2024}.
    \item \textsc{Log Sampling} --- Sample and prioritize logs to reduce the amount of data that needs to be stored and processed~\cite{Albuquerque-2024}.
    \item \textsc{Distributed Tracing} --- Assign each external request a unique ID and record how it flows through the system from one service to the next in a centralised server that provides visualisation and analysis, making troubleshooting the application faster and less complicated (see Page~\ref{pat:distributed-tracing}).
    \item \textsc{Deployment Tracking} --- Track every deployment and change to the production environment, making it possible to relate effects observed in the system to changes that caused them~\cite{Albuquerque-2023}.
    \item \textsc{Exception Tracking} --- Send exceptions to a centralised exception tracking service that aggregates exceptions, tracks their resolution, and creates alerts~\cite{Albuquerque-2023}.
    \item \textsc{Liveness Endpoint} --- Implement a specialised endpoint that responds to requests without side effects. Then, configure another system (\textit{e.g.}, service, tool, load balancer) to periodically check that endpoint and take action when that fails, providing an automatic way to detect that the instance is unable to respond~\cite{Albuquerque-2022}.
    \item \textsc{Readiness Endpoint} --- Implement a specialised endpoint that checks if the service is ready to accept and process traffic. Then, configure another system (\textit{e.g.}, service, tool, load balancer) to periodically check that endpoint and stop routing traffic to the service when the check fails~\cite{Albuquerque-2022}.
    \item \textsc{Synthetic Testing} --- Create or pick a subset of existing test cases and periodically run them against the production environment, ensuring the application behaves as expected and detecting issues before they affect end-users~\cite{Albuquerque-2022}.
    \item \textsc{Application Metrics} --- Instrument the application to gather business and performance metrics. Collect these metrics in a centralised service that provides aggregation and visualisation, allowing deeper insight into the application's performance (see Page~\ref{pat:application-metrics}).
    \item \textsc{Infrastructure Metrics} --- Instrument the server and runtimes to capture relevant metrics of the operative system and underlying infrastructure and collect them in a centralised server, allowing the team to get a real-time overview of the application's environment (see Page~\ref{pat:infrastructure-metrics}).
\end{enumerate}
}

The patterns in this paper are \textsc{Distributed tracing}, \textsc{Application Metrics}, and \textsc{Infrastructure Metrics}. We describe each of them using this structure:

{\small
\begin{itemize}
    \item \textbf{Name} --- intuitive or established name for the pattern.
    \item \textbf{Summary} --- short summary of the pattern, focusing on the core of its problem and solution.
    \item \textbf{Context} --- contextualization of the pattern; provides background on the problem and may also refer to other design patterns that can be considered before the current one.
    \item \textbf{Problem} --- a brief description of the problem as a question.
    \item \textbf{Forces} --- a list of forces constraining the solution to point in a certain way and not another.
    \item \textbf{Solution} --- starts with a sentence in \textit{italics} that captures the gist of the solution and goes on to describe the solution for the problem, where we highlight in \textbf{bold} certain keywords that represent the roles of different components or modules; these components and their interactions are then depicted in a figure by the end of the solution section.
    \item \textbf{Consequences} --- a bullet-point description of the pattern's main advantages and disadvantages that should be considered when adopting it.
    \item \textbf{Example} --- an illustrative example, real or fictional, of the pattern in use.
    \item \textbf{Known Uses} --- succinct description of real-world cases that use the pattern; throughout this section, we also briefly mention existing tools that can be used to adopt the pattern.
    \item \textbf{Related Patterns} --- a few paragraphs describing how the pattern relates to other design patterns, both from the catalogue and other works.
\end{itemize}
}

\section{Distributed Tracing}\label{pat:distributed-tracing}
Cloud applications often involve multiple modules working together to handle user requests, making it challenging to pinpoint failures or performance bottlenecks across nested operations. Therefore, assign each external request a unique ID and trace its flow through the system in a centralised server, to enable visualisation and analysis, simplify troubleshooting and improve system observability.

\subsection{Context}
Cloud applications are usually a sum of modules that work together to reply to user requests. Therefore, a request may require many nested operations throughout the application's modules, both internal and external. When something breaks during these operations, it can become hard to figure out exactly where the failure is. Moreover, the person troubleshooting the error may not even be familiar with all the modules, making the troubleshooting process longer and more complicated.

Sometimes, even if the system did not fail, it might still be underperforming. If a user needs to wait five or more minutes for a webpage to load, there might be an operation (among all executed for that request) that is taking too long. The cause of slowness may be hard to pinpoint when requests are complex and involve many operations.

\subsection{Problem}
How can developers record and visualise the end-to-end behaviour of the application to identify problems during the request lifetime?

\subsection{Forces}
Addressing this problem is subject to the following forces:

\begin{itemize}
    \item Metrics can provide an aggregate view of the system (\textit{e.g.}, max response time in the last 15 minutes, the sum of request errors in the last hour), but they are not helpful when trying to understand the behavior throughout the system induced by a particular request.
    \item Logs can capture detailed information for specific requests, but they are often scattered across different services and infrastructures, making correlation and analysis time-consuming.
    \item Logs usually record events or errors, but they do not provide clear visibility into the time each operation took, to help identify bottlenecks.
    \item Synthetic tests~\cite{albuquerque2022proactive} and unit tests can simulate requests and catch some issues, but they can easily miss transient runtime conditions that may surface in production such as spikes in resource usage or network issues.
    \item Recording detailed trace data can help diagnose issues, but it must be done in a way that minimizes the impact on system performance and maintainability.
    \item The different services of cloud-native applications often need to collaborate to fulfill a request, but the complexity of inter-service dependencies makes it difficult to understand and debug the system’s behavior.
    \item Cloud-native applications often run across heterogeneous environments---using multiple programming languages, frameworks, and infrastructures---which increases the effort of instrumenting different services.
\end{itemize}

\subsection{Solution}
\textit{Assign each external request a unique ID and record how it flows through the system from one service to the next in a centralised server that provides visualisation and analysis, making troubleshooting the application faster and less complicated.}
\vspace{0.3cm}

\begin{figure}[b]
    \centering
    \includegraphics[width=\textwidth]{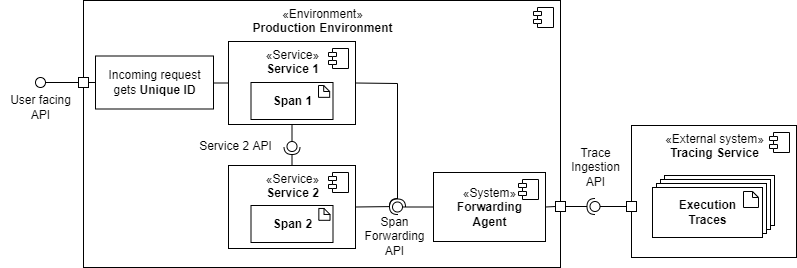}
    \caption{Overview of the structure for the \textsc{Distributed Tracing} pattern. The request is assigned a unique ID as it enters the system and the generated spans carry that ID.}
    \label{fig:distributed-tracing}
\end{figure}

Namely, instrument the source code as follows:
\begin{itemize}
    \item An \textbf{Unique ID} is assigned to each incoming request (see \textsc{Correlation ID}~\cite{Brown-Woolf-2016});
    \item The ID is sent to every \textbf{Service} that processes the request;
    \item Local thread activity is recorded in a \textbf{Span} of execution and tagged with the unique ID;
    \item The unique ID is included in all the logs generated while processing the request;
    \item The span information is sent to a central \textbf{Tracing Service} directly from the microservice or through a local \textbf{Forwarding Agent} (see Figure~\ref{fig:distributed-tracing}).
\end{itemize}

The collector is then responsible for processing all spans and constructing \textbf{Execution Traces} by mashing together the related spans and logs. Usually, the collector also provides a UI to visualise request information and the ability to query the traces. The collector lets developers analyse the whole execution trace and quickly figure out performance bottlenecks, which operation originated a failure and the logs associated with the previous situations.

Because cloud-native systems can change quickly---due to autoscaling, rolling updates, or container restarts---the infrastructure and service instances that generate a trace may no longer exist when a developer investigates an issue. To address this, spans can include metadata about the infrastructure and service (\textit{e.g.}, service version, instance ID, container name, deployment timestamp, or node label) so that it is possible to understand what the general operational context was when later trying to interpret traces. When using \textsc{Deployment Tracking} or \textsc{Infrastructure Metrics}, the data they generate can also be used to understand the system state at the time of execution, and often be correlated with traces through the same metadata.

When recording local thread activity in the spans, developers should consider following the OpenTelemetry API, which has seen growing adoption in the software industry~\cite{Newman-2021}. Particularly in heterogeneous environments, implementing distributed tracing can become more complex due to the varying levels of support for instrumentation. The language-specific SDKs made available by OpenTelemetry with support for multiple platforms (\textit{e.g.}, Java, Python, Go, Node.js) can reduce this burden.

When adopting this pattern, we must consider how to handle the information collected, as collecting and storing traces for every request may come to have a negative impact on the system. A way to address this concern is tail-based sampling, \textit{i.e.} sampling the execution traces after they have been reconstructed in the collector service. The team can decide to drop some of the traces and only store the remaining ones, effectively reducing the overhead on the system. \textit{``The idea here is to capture enough information to understand what our system is doing, but not capture so much information that the system itself cannot cope''}~\cite{Newman-2021}. The sampling can be random or dynamic. The former consists of randomly selecting which traces to sample, increasing the likelihood of losing relevant information or collecting insufficient information to reach relevant conclusions. The latter has different sampling rates that depend on certain conditions (\textit{e.g.}, if there was an error during execution or if a given operation has been slower than usual). Dynamic sampling is harder to implement and requires a bit more processing, but it is more likely to catch the information needed to troubleshoot the application. Balancing this trade-off is the responsibility of the team when adopting this pattern.

When adopting this pattern, we must also consider how to balance the benefit of detailed observability with its costs. Tracing every request introduces overhead, both in runtime performance and in the resource consumption of the tracing infrastructure. To mitigate this, teams often adopt sampling strategies to reduce the volume of traces stored and analyzed while preserving the utility of traces for debugging and performance analysis. \textit{Tail-based} sampling makes the decision to retain a trace after all its spans are complete, often based on criteria depending on information of the spans themselves; unlike \textit{head-based} sampling, which filters at the start of a request, usually randomly. 

Another important concern is time synchronization across services, especially in distributed infrastructures where clock skew can occur due to unsynchronized or faulty NTP configurations. This can result in incorrect ordering and inaccurate duration of spans, leading to misleading trace visualizations. To address this, teams should ensure that all nodes use a reliable time synchronization service, such as NTP, and leverage clock skew adjustment mechanisms provided by tracing tools. Additionally, to ensure consistent span ordering regardless of clock discrepancies, each span should include the ID of its parent span. This hierarchical structure allows the tracing system to reconstruct the correct sequence of operations, even when physical clocks are out of sync. Tools like Jaeger\footnote{More details about Jaeger are available at \href{https://www.jaegertracing.io/}{jaegertracing.io}.}, which support parent span IDs and built-in clock skew adjustment, can simplify the implementation of this approach and improve the accuracy of distributed traces.

\subsection{Consequences}
This pattern has the following advantages:
\begin{itemize}
    \item The mean time to detect (MTTD) becomes shorter, making troubleshooting the system faster.
    \item User requests become observable from end to end, \textit{i.e.} every operation performed since the request first arrives in the system until a response is sent to the user is tracked.
    \item Traces provide insight into individual operations, making it easier to identify system bottlenecks.
\end{itemize}

\noindent This pattern also has the following drawbacks:
\begin{itemize}
    \item Depending on the existing library support when instrumenting code in a particular programming language used, implementing this pattern may require changes that imply a non-trivial effort.
    \item Instrumenting the source of different services and aggregating traces introduces overheads that may degrade performance, increase latency, and raise complexity in code maintenance.     
    \item Traces need to be stored, and as time goes by, they will start to occupy more and more space, demanding considerable infrastructure and increasing the storage costs.
    \item Since the solution is a bit complex, its learning curve may create some initial friction with the team.
\end{itemize}

\subsection{Example}
Meesho is an online shopping app for India. A blog post by Agarwal~\cite{Agarwal-2022} explains that, as they transitioned from a monolithic architecture to a microservices-based framework, understanding the path of a single request became much more complicated. Because \textit{"a single request passes through multiple services"}~\cite{Agarwal-2022}, tracing a request across all services became a challenge. The author explains in further detail their feed system's architecture. Most of their services composing the feed system were \textit{"Spring framework-based Java microservices"}~\cite{Agarwal-2022}. It followed a layered architecture, \textit{"where each layer has its responsibility and each domain service operates within its boundary context"}~\cite{Agarwal-2022}. 

To tackle the abovementioned issue, the team decided to implement distributed tracing in their system through \textit{Spring Cloud Sleuth}. The author states that the decision was based on the tool's \textit{"auto-configuration capability and compatibility with other Spring libraries"}~\cite{Agarwal-2022}. The distributed tracing works as follows:
\begin{itemize}
    \item The unique IDs are initialized by the first service that does not find them in the request headers. They use two unique IDs for each request---the \textit{Trace ID} and the \textit{Span ID}---and the service attaches them to \texttt{ThreadLocal}, a Java class that provides thread-local variables.
    \item The team \textit{"implemented an interceptor that passes these IDs downstream as headers"}~\cite{Agarwal-2022}.
    \item To include the unique IDs in the logs, the team used SLF4J to fetch the IDs from the \textit{ThreadLocal} variables and include them in the logs.
\end{itemize}

They bumped into a problem with asynchronous processing that led to the loss of both unique IDs because the thread context changed in the middle of the execution. To maintain trace context across threads, the team manually propagated \textit{ThreadLocal} variables containing the trace and span IDs. This often involved wrapping asynchronous tasks (\textit{e.g.}, \textit{Runnable}, \textit{Callable}) so that the trace context was restored before execution. This workaround ensured trace continuity during async processing. With that out of the way, Sleuth handles the aggregation and correlation of spans to construct the execution traces.

The author ends by explaining that to diagnose an issue reported by a particular user they queried trace logs using a User ID to locate relevant Trace IDs---\textit{"This showed us the whole request trace across services, and we instantly figured out that the A/B service was returning an unexpected response in one of our domain services which caused that issue"}~\cite{Agarwal-2022}.

\subsection{Known uses}
Agarwal's~\cite{Agarwal-2022} example described in the previous section is already a known use of this pattern that utilised Spring Cloud Sleuth to implement this pattern in a real-world microservices architecture.

FloQast is another company that uses distributed tracing to understand its application. In a blog post, Dinh~\cite{Dinh-2020} explains that FloQast is a fast-growing company, so they \textit{"add more and more business logic to the codebase every single day"}~\cite{Dinh-2020}. The team noticed that bottlenecks started to appear in the system, but they did not have enough information to follow a data-driven decision. Thus, they started using AWS X-Ray service, which, as the author explains, \textit{"allows users to gain insight into requests that your application serves"}~\cite{Dinh-2020}. Using that tool, they can \textit{"easily trace requests across AWS resources and other microservices"}~\cite{Dinh-2020}.

Finally, and on a much bigger scale, Netflix built its distributed tracing infrastructure to solve user issues better. In Netflix Technology Blog, Pandey~\cite{Pandey-Netflix-2020} goes into some detail about how they did and evolved the infrastructure to cope with emerging standards like Open-Zipkin and Open-Tracing. \textit{"Investigating a video streaming failure consists of inspecting all aspects of a member account"}~\cite{Pandey-Netflix-2020}, so troubleshooting a user complaint can become very complicated on a system as complex as Netflix. They developed Edgar, an internal tool for troubleshooting streaming sessions. They started with a simple tool that, based on unique IDs for each streaming session, was able to \textit{"reconstruct session failure by providing service topology, retry and error tags, and latency measurements for all service calls"}~\cite{Pandey-Netflix-2020}. For their case in specific, the team had to apply a hybrid head-based sampling approach that enabled them \textit{"to record 100\% traces in our mission-critical streaming microservices while collecting minimal traces from auxiliary systems like offline batch data processing"}~\cite{Pandey-Netflix-2020}. Even though this helped reduce the number of traces stored, the author states that they still had issues escalating their ElasticSearch clusters. So they eventually transitioned to Cassandra clusters to handle the high data ingestion rates.

\subsection{Related patterns}
The \textsc{Correlation ID}~\cite{Brown-Woolf-2016} pattern provides a basic mechanism for correlating logs across services by assigning a unique identifier to each request. While this can help trace the flow of a request through the system, it does not provide detailed timing information or visualizations of the request's lifecycle. \textsc{Distributed Tracing} builds on this by offering a more comprehensive solution, including the ability to measure latency, identify bottlenecks, and visualize the sequence of operations across services. Teams can start with \textsc{Correlation ID} to address basic correlation needs and later adopt \textsc{Distributed Tracing} for deeper insights and more troubleshooting capabilities.

Logs are a common way to implement execution spans. In such cases, adopting \textsc{Log Aggregation}~\cite{sousa2017engineering,sousa2025pattern} becomes almost necessary to get \textsc{Distributed Tracing}. In addition, logs can also be correlated and aggregated with the traces of execution. This provides the team with a complete report of the events and the time each operation took across the whole request lifetime.

However, generating too many traces can carry a very high overhead. The team should consider the \textsc{Log Sampling} pattern to implement a sampling strategy that minimises the overhead of logged traces while maintaining the ones that truly matter for troubleshooting purposes.

\section{Application Metrics}\label{pat:application-metrics} 

Cloud abstracts infrastructure details, making the concerns about application performance more about user responsiveness than resource efficiency. Understanding performance and usage patterns is vital for maintaining quality. Therefore, instrument the application to collect business and performance metrics, aggregating and visualizing them in a central service to gain deeper insights into the application's behavior.

This pattern is sometimes also known as Inside-out Health Check~\cite{Gupta-Patterns-2022} or Monitoring Metrics~\cite{Waseem-Practices-2021}.

\subsection{Context}
Users expect software systems to bring value through appealing features, but also for it to fail rarely and work with acceptable speed, which pushes teams to keep track of their system’s performance to ensure it meets certain requirements. These may be informal and more flexible, such as an end-user of a web application that expects a page to load in less than 30 seconds, or formal and strict, such as a service level agreement (SLAs) between an infrastructure as a service (IaaS) provider and its client. Both cases are equally important for a company since it might be losing customers or having to compensate them for a contractual breach, respectively.

Since the cloud abstracts away the infrastructure on which the code is running, performance becomes more about how well the system can respond to the user than how efficiently it uses the available resources. Moreover, performance can be seen from the technical and business lenses. Understanding customer trends and satisfaction is just as important as knowing availability and the number of errors, for example. Both can bring helpful insight to a company and give it a more significant market presence.

\subsection{Problem}
How can the team get an overview of how the application is performing and understand the emerging usage patterns?

\subsection{Forces}
Addressing this problem is subject to the following forces:

\begin{itemize}
    \item Logs can be used to to extract relevant system performance measures, but processing them is computationally expensive and does not provide real-time visibility.
    \item Simulating requests and testing in production can surface issues in predefined scenarios, but they offer limited data and may miss unexpected or new conditions.
    \item Infrastructure (\textit{e.g.}, cloud) providers often supply infrastructure metrics, but these may lack application-specific detail needed for fine-grained monitoring.
    \item Reusable monitoring solutions are desirable for reducing cost and effort, but they still need to be adapted to the specific needs of diverse applications.    
    \item Cloud-native applications often run across heterogeneous environments---using multiple programming languages, frameworks, and infrastructures---which increase the effort of instrumenting different services.    
\end{itemize}

\subsection{Solution}
\textit{Instrument the application to gather business and performance metrics. Collect these metrics in a centralised service that provides aggregation and visualisation, allowing deeper insight into the application's performance.}
\vspace{0.3cm}

\begin{figure}[b]
    \centering
    \includegraphics[width=0.75\textwidth]{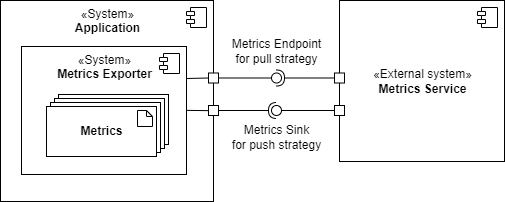}
    \caption{Overview of the structure for the \textsc{Application Metrics} pattern. The solutions usually only follows the pull or the push strategy. We represented both for completeness.}
    \label{fig:application-metrics}
\end{figure}

Namely, the team can collect \textbf{Metrics} on many things, such as request rate, error rate, customer orders, and user posts. Ideally, one would measure everything, but that introduces more overhead and requires considerable infrastructure. That said, the first decision the team must make when adopting this pattern is to define which \textbf{Application} metrics they wish to monitor (see Figure~\ref{fig:application-metrics}). Consider The Four Golden Signals proposed in~\cite{Ewaschuk-SRE-2016} as a starting point. They are briefly presented below, but for further detail refer to the original work:
\begin{itemize}
    \item Latency - the time between when a user request arrives in the system, and a response is issued to the user.
    \item Traffic - a \textit{"high-level system-specific metric"}~\cite{Ewaschuk-SRE-2016} that measures the number of incoming requests in your system.
    \item Errors - the rate of requests that result in a failure, of any kind, by a period of time.
    \item Saturation - the percentage of system resources being utilised at the moment.
\end{itemize}

Other guidelines can also be used, like the RED method~\cite{Jackson-Red-2018} (rate of requests, errors, duration) and the READS metrics~\cite{Gupta-READS-2022} (rate of requests, errors, availability, duration/latency, and saturation), and the team should consider which one is more suited to their context. Typically, a team will adopt RED where there is a large population of homogeneous microservices and minimising per-endpoint overhead is a priority, and READS when it is relevant to distinguish availability from raw error rates and to incorporate resource pressure into health assessment. 

Beyond service/runtime indicators, teams also track business (domain) metrics that show whether the product is delivering value (\textit{e.g.}, orders placed, successful checkouts, active users, trial‑to‑paid conversion, average order value, refund or failure rates, feature adoption). These should be few, purposeful KPIs tied to goals, as the more we collect, the harder it becomes to process and derive useful insights. 

Another thing to consider is the data resolution---\textit{i.e.} in this case, the number of data points collected by time period. Depending on the current context of the system, some metrics may be collected and stored in a greater resolution than others. As Newman exemplifies: 

\begin{quote}
    \textit{"I might want a CPU sample for my servers at the resolution of one sample every 10 seconds for the last 30 minutes, in order to better react to a situation that is currently unfolding. On the other hand, the CPU samples from my servers from last month are likely needed only for general trend analysis, so I might be happy with calculating an average CPU sample on a per-hour basis." }~\cite[p.~322]{Newman-2021}
\end{quote}

Therefore, when adopting this pattern, the team should consider having the ability to change the resolution of the data that is stored or being collected. That allows the team to better manage storage space and overhead while still collecting a sufficient number of metric points for troubleshooting.

Another important consideration is granularity. Fine-grained metrics (\textit{e.g.}, request duration per user per endpoint) provide rich insights but will increase storage and processing overhead. Coarse-grained metrics (\textit{e.g.}, average request duration per service) reduce overhead but may hide anomalies. The team can balance this trade-off based in the power of each metric to provide insights and the criticality of the functionality---\textit{e.g.}, detailed metrics might be necessary for payment processing endpoints, while summary metrics may be enough for background jobs.

Once we know \textit{what} to measure, we should understand \textit{how} to do it. Each metric is sampled periodically by a \textbf{Metrics Exporter} (see Figure~\ref{fig:application-metrics}) and is usually composed of a name, value and timestamp~\cite{Richardson-Wiki-2021}. Further context must be added to the metric through tags, \textit{i.e.} key-value pairs for external properties, that are collected each time step. The higher the cardinality of the added tags (\textit{e.g.}, usernames and emails, which are unbounded sets), the more the metrics system will struggle to keep all the information. When adopting this pattern, the team should consider reducing the cardinality of these tags or using a system that is prepared to receive high cardinality data.

Some of these tags will be related to the runtime services and infrastructure where the metric originated. This allows metrics to be grouped and queried in context of the system structure at the time the metric was generated, even if that structure has changed as, in cloud environments, service instances may have been replaced or terminated before developers even feel the need to examine their metrics. Correlating this metadata with data generated by \textsc{Deployment Tracking} can also help to reconstruct the resource topology at the time of an incident.

Finally, metrics generated and spread across many services are not very helpful. To make actual use of them, the team needs a \textbf{Metrics Service} - a centralised system to collect and aggregate (\textit{e.g.}, to calculate the average, sum or percentiles) the metrics and present them in a friendly way to the team (see Figure~\ref{fig:application-metrics}). This can be achieved in two ways~\cite{Richardson-Book-2018}:
\begin{itemize}
    \item push - the application sends the metrics to an API provided by the metrics service (\textit{e.g.}, AWS CloudWatch).
    \item pull - the metrics service fetches the metrics data from an API provided by the application (\textit{e.g.}, Prometheus).
\end{itemize}

Both options are viable; it all depends on the context of the system. Usually, the pull model is more transparent to the application. In other words, it does not require as much instrumentation, keeping the business logic more isolated from these monitoring details.

\subsection{Consequences}
This pattern has the following advantages:
\begin{itemize}
    \item It provides an up-to-date overview of the application's state.
    \item It provide valuable insights from both technical (\textit{e.g.}, capacity planning, predicting potential problems) and business (\textit{e.g.}, strategic decision-making, understanding trends) perspectives, enabling data-driven decisions.
    \item Metrics enable the team to notice non-evident patterns and problems that would most likely go unnoticed otherwise, when paired with alerts, act before they ffect the users and the business.
\end{itemize}

\noindent This pattern also has the following drawbacks:
\begin{itemize}
    \item Aggregating and storing a high number of metrics may imply considerable infrastructure costs.
    \item Instrumenting the source of different services introduces overheads that may degrade throughput and latency, and raise complexity in code maintenance.     
    \item Configuring the metrics themselves, including deciding what to measure, setting thresholds, and managing granularity, can require significant effort and expertise.
    \item Collecting application data in the cloud depends heavily on what the team has access to, so this pattern gets harder to adopt the fewer access rights the team has to the deployment environments.
\end{itemize}

\subsection{Example}

A post by Campuzano on the GumGum Tech Blog~\cite{Campuzano-2019} explains how the company adopted Prometheus to monitor its systems. Before the company adopted the microservice architecture, the author states that monitoring was quite simple. Their \textit{"applications, servers, and services were pretty much fixed and well-known"}. However, things got much more complicated once they started containerising their applications and adopting the microservices architecture---\textit{"the explosion in the quantity and complexity of the systems to be monitored was neither manageable nor sustainable with the legacy monitoring stack that we had in place"}.

\begin{figure}[!ht]
\centering
\includegraphics[width=\textwidth]{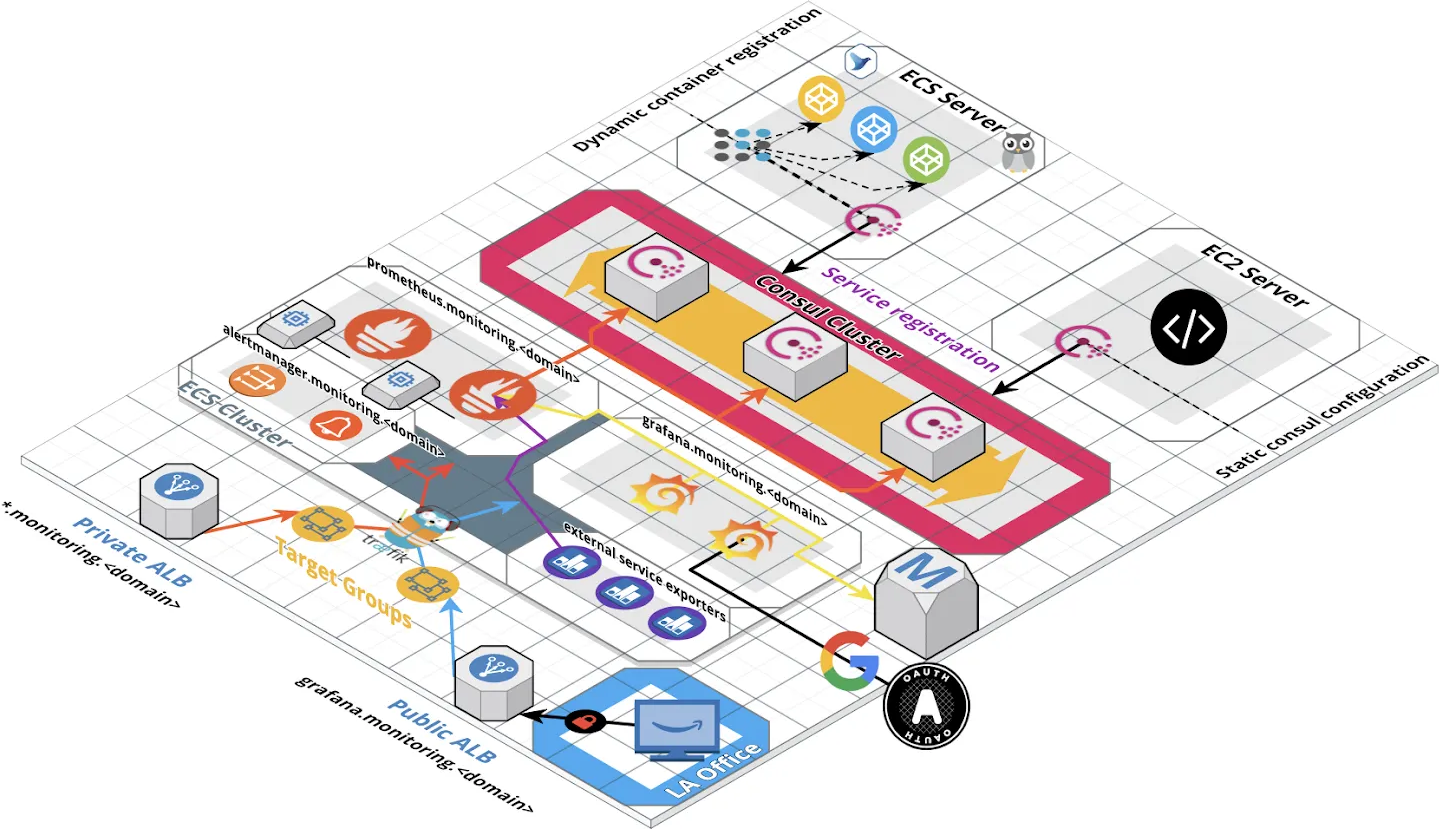}
\caption{Real-world example of the \textsc{Application Metrics} pattern~\cite{Campuzano-2019}.}
\label{fig:application-metrics-example}
\end{figure}

To address the new monitoring needs, they used Prometheus to create a monitoring stack capable of handling the complexity. According to the author, they \textit{"chose Prometheus among other systems, mostly because of its flexibility, extensibility and huge open source community backing the project"}. They used different Prometheus exporters to instrument the code to generate metrics. The tool follows a \textit{pull} strategy to get the metrics, \textit{i.e.} it periodically scrapes them from a configurable endpoint. GumGum's monitoring stack collected metrics from the application, containers and servers. Additionally, they used Grafana to visualise the collected metrics on customised dashboards and Alert Manager to configure and manage alerts on top of the Prometheus metrics. Figure~\ref{fig:application-metrics-example} is part of the post and depicts the overall Prometheus architecture used by GumGum.

The author does not precisely mention which metrics they collect, but these could include the number of ad impressions per second, rendering latency per ad unit, failed ad fetches per service, and average user session duration. Tracking these metrics could enable regional scaling policies and improving ad display success rate. Additionally, Campuzano mentions that they used Prometheus Pushgateway for \textit{"ephemeral and batch jobs to expose its metrics to Prometheus"}~\cite{Campuzano-2019}. These jobs send the metrics to the Pushgateway, which Prometheus then scrapes. Thus, they could collect application-level metrics and visualise them in specific dashboards for their teams. With Prometheus and Grafana, GumGum became able to collect metrics from many parts of their system and \textit{"create beautiful and meaningful Grafana dashboards with those metrics"}~\cite{Campuzano-2019}.

\subsection{Known uses}
Campuzano's~\cite{Campuzano-2019} example described in the previous section is already a known use of this pattern that used Prometheus and Grafana to collect metrics from a microservice architecture.

A post by Domenico Stragliotto~\cite{Stragliotto-2019}, a backend developer for THRON, elaborates how the company adopted metrics on their system. The author explains they were looking to improve their ability to analyse long-term trends, build dashboards, troubleshoot and get alerted. To achieve their goals, they used Prometheus to collect metrics from their services. They followed the RED method~\cite{Jackson-Red-2018}, so, for each service, they collected the number of requests per second, the number of errors per second and the amount of time to process each request. These metrics were then exposed in Grafana dashboards so the teams could easily view each service's state.

Finally, an infrastructure software engineer post on callstats.io's blog~\cite{Callstats-2018} reports how the company uses Prometheus to monitor its services. They chose Prometheus due to its seamless integration with Kubernetes, powerful query language and operational simplicity. The engineer states that the company's developers \textit{"use Prometheus to compare performance and resource usage between service releases"}~\cite{Callstats-2018}, while the analytics team uses it \textit{"for several artificial intelligence-related metrics"}~\cite{Callstats-2018}.

\subsection{Related patterns}
Although the cloud abstracts the infrastructure where the application runs, understanding the performance of the host machine can be helpful. The underlying machine's performance can impact the application's performance. Suppose the team needs a more comprehensive view of the system or further correlation between the application and its infrastructure. In that case, they should consider adopting~\textsc{Infrastructure Metrics} to complement the metrics provided by this pattern. Additionally, patterns like \textsc{Liveness Endpoint}, \textsc{Readiness Endpoint}, and \textsc{Synthetic Testing} can supply valuable metrics that provide further insights into its behavior and health.

\section{Infrastructure Metrics}\label{pat:infrastructure-metrics} 
Software relies on infrastructure, and resource scarcity like CPU, memory, or disk can lead to slowness or outages. Teams need visibility into their application's infrastructure to diagnose and correlate issues effectively. Therefore, instrument servers and runtimes to capture key metrics of the operating system and infrastructure, and centralize this data for real-time monitoring and analysis.

This pattern is sometimes also known as Inside-Out Health Check~\cite{Gupta-Patterns-2022}.

\subsection{Context}
Software needs infrastructure to execute. Even though the cloud abstracts this infrastructure from the developing team, applications still depend on the underlying resources. Scarcity of resources like CPU, memory or disk can bring extreme system slowness or even an outage. In addition, these issues can be caused by faults in the code (\textit{e.g.}, memory leaks), so it is the developing team's responsibility to find and fix them. The team should monitor the essential infrastructure resources to identify code-level problems and correlate end-user issues with the infrastructure.

In a pay-as-you-go scenario, which allows for greater flexibility and cost savings, it will drive the costs up if the application starts to utilise more and more resources. Being able to notice the increasing usage of infrastructure is necessary so the team can figure out if that is due to greater demand on the application (\textit{e.g.}, a new product release with a significant influx of users), in which case it is to be expected, or an issue with the code (\textit{e.g.}, poorly optimised loops or memory leaks) that needs to be fixed. Moreover, public clouds are usually under service level agreements (SLAs) that declare the minimum requirements the infrastructure must provide. If a given resource constantly drops below the agreed threshold, the cloud consumer should be able to confront the provider and take action according to what is stated in the SLA. Thus, understanding the performance and state of the infrastructure is relevant for both cloud providers and consumers, so the machines the software is running on should be monitored.

\subsection{Problem}
How can the team know the state of their application's infrastructure and correlate it with ongoing problems?

\subsection{Forces}
Addressing this problem is subject to the following forces:

\begin{itemize}
    \item The system must scale when resource thresholds are crossed, but some tools and cloud providers require manual intervention, which can slow response time.
    \item Cloud environments support flexible resource usage, but uncontrolled consumption can lead to unexpectedly high operational costs.
    \item Application services may be spread across different infrastructure resources, but this distribution complicates centralized monitoring and analysis.
    \item Resource usage can fluctuate significantly with load, but provisioning and scaling decisions often rely on static resource thresholds that may not adapt quickly enough.
    \item Infrastructure provisioning is often governed by contractual SLAs, but teams may lack the tooling or access needed to verify compliance in real-time.
    \item Reusable monitoring solutions can help reduce development effort, but they still need to be adapted to the specific infrastructure needs of applications.
    \item Cloud-native applications often run across heterogeneous environments---using multiple programming languages, frameworks, and infrastructures---which increase the effort of instrumenting different services.    
\end{itemize}

\subsection{Solution}
\textit{Instrument the server and runtimes to capture relevant metrics of the operative system and underlying infrastructure and collect them in a centralised server, allowing the team to get a real-time overview of the application's environment.}
\vspace{0.3cm}

As explained in Application Metrics (see Section~\ref{pat:application-metrics}), \textbf{Metrics} are usually composed of a name, a value, and a timestamp~\cite{Richardson-Book-2018}. Tags can be stuffed into the metrics to provide further context. Tags are key-value pairs of properties, so they can come with nearly any kind of information. However, the higher the number of different values each tag can have, or, in other words, the higher its cardinality, the harder it is to store and process.

Since infrastructure components like virtual machines, containers, and disks can be provisioned and decommissioned dynamically, it is often useful for exporters or monitoring agents to also associate tags with infrastructure metrics that can provide context of what the infrastructure was like at the time the metrics were collected. This is particularly important when diagnosing issues long after they occurred, as the relevant infrastructure may no longer exist. Correlating this metadata with data generated by \textsc{Deployment Tracking} can also help to reconstruct the resource topology at the time of an incident.

\begin{figure}[h]
    \centering
    \includegraphics[width=0.75\textwidth]{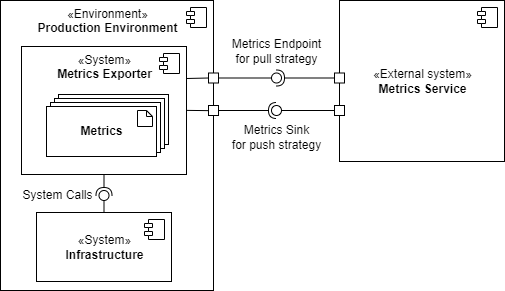}
    \caption{Overview of the structure for the \textsc{Infrastructure Metrics} pattern. The solutions usually only follow the pull or the push strategy. We represented both for completeness.}
    \label{fig:infrastructure-metrics}
\end{figure}

Instrumentation can take many forms, but it is essentially the act of adding measuring instruments to the system. These instruments collect the \textbf{Infrastructure} information and generate metrics out of it. In the context of this pattern, instrumenting the server and runtimes means having a \textbf{Metrics Exporter} that is responsible for gathering a predefined set of metrics (see Figure~\ref{fig:infrastructure-metrics}). It must have access to the OS and underlying infrastructure and, ideally, be configurable to specify what should be captured and at what resolution (\textit{i.e.}, the number of data points per time interval).

Some public clouds, like GCP, provide an API with each application engine that exposes infrastructure metrics. Note that this API, by itself, is already a suitable exporter. It is gathering the necessary information and exposing it. Even though this kind of API may not be configurable (\textit{e.g.}, you can not change the sampling rate), the team can request exclusively the metrics they need and at a larger interval than what the exporter samples.

Configuring the sampling rate and granularity may be necessary since storing and processing large amounts of data can demand considerable infrastructure and slow the overall monitoring process. Therefore, when adopting this pattern, the team should decide how many data points should be collected and how many should be stored. They should also decide on the level of detail captured over time---for instance, per-second CPU usage versus per-minute averages. Fine-grained metrics allow detailed analysis of short-term behavior, but they consume more resources and can lead to data bloat. In contrast, coarse-grained metrics reduce overhead but may hide short-lived issues. Teams should balance this trade-off based on operational needs, criticality of the metric, and cost.

Another crucial decision is what metrics to monitor. As previously stated, the more metrics the team collects, the easier it is to understand the infrastructure's state. Thus, the team should start with the most useful ones for their case and expand as the need arises. The USE method is a common method to help with this decision~\cite{Gregg-2013}. It suggests that the following metrics should be the first to be collected:

\begin{itemize}
    \item Utilisation - \textit{"the percentage of time that the resource is busy servicing work during a specific time interval"}~\cite{Gregg-2013};
    \item Saturation - \textit{"the degree to which the resource has extra work which it cannot service, often queued"}~\cite{Gregg-site-2017};
    \item Errors - \textit{"the count of error events"}~\cite{Gregg-2013}.
\end{itemize}

Note that the USE method is more than just an indication of the metrics that should be gathered. It also incentivises the team to start by identifying the resources in their system (\textit{e.g.}, CPU, memory, storage). Only then should they define what each of the above metrics means to each resource and how they should be gathered. That way, the USE method increases the overall understanding of the system---when the team notices a missing metric that would be useful to solve a currently ongoing application problem, they can start from the list of metrics they know they are not gathering.

Infrastructure metrics play an important role in managing elasticity. Autoscaling is often based on thresholds defined for specific metrics, such as CPU or memory use, with these metrics acting as triggers for reactive scaling (\textit{e.g.}, scaling out when usage surpasses 80\%). Teams can also proactively scale to prevent performance degradation, based on the prediction of traffic surges from historical patterns in infrastructure metrics. Scaling decisions have cost implications, with more instances meaning more compute charges. The choice of metrics to be monitored and associated scaling policies can be used to appropriately balance performance and cost. \textit{E.g.}, frequent CPU spikes might call for load distribution optimization rather than scaling up resources.

Finally, metrics generated and spread across many services are not very helpful. To make actual use of them, the team needs a \textbf{Metrics Service} - a centralised system to collect and aggregate (\textit{e.g.}, calculate the average, sum, or percentiles) the metrics and present them in a friendly way to the team (see Figure~\ref{fig:infrastructure-metrics}). This can be achieved in two ways~\cite{Richardson-Book-2018}:
\begin{itemize}
    \item push - the application sends the metrics to an API provided by the metrics service (\textit{e.g.}, AWS CloudWatch);
    \item pull - the metrics service fetches the metrics data from an API provided by the infrastructure (\textit{e.g.}, Prometheus)
\end{itemize}

Both options are viable. It all depends on the system's context. Usually, the pull model is more transparent to the application. In other words, it does not require as much instrumentation, keeping the business logic more isolated from these monitoring details.

\subsection{Consequences}
This pattern has the following advantages:
\begin{itemize}
    \item Metrics enable the team to notice non-evident patterns and problems that would most likely go unnoticed otherwise.
    \item It provides a real-time and straightforward overview of the infrastructure's state.
    \item Operations teams can more easily cut down on unused resources, decreasing the system's costs.
    \item Issues with the code that directly impact the infrastructure (\textit{e.g.}, memory leaks) become easier to detect.
    \item It allows for correlation between user symptoms, like recurring failures when creating a new post, to infrastructure problems, like a full database.
\end{itemize}

\noindent This pattern also has the following drawbacks:
\begin{itemize}
    \item Aggregating and storing a high number of metrics may require considerable infrastructure, increasing its costs.
    \item Instrumenting the server and runtimes to collect the metrics introduces overheads that may degrade performance, increase latency, and raise complexity in code maintenance.     
    \item Correlating infrastructure and application metrics is based on the time of events; since the systems collecting the metrics are different, their clock times may also diverge slightly, making way for wrong relations between both metric types.
    \item Data can quickly become separated into different categories (or silos) without actual correlation, which heavily reduces this pattern's benefits.
    \item Collecting infrastructure data in the cloud depends heavily on what the team has access to, so this pattern gets harder to adopt the fewer access rights the team has to the infrastructure.
    \item In PaaS and FaaS scenarios, the only way to collect infrastructure metrics may be through the cloud provider's dedicated solutions, which reduces the team's flexibility and restricts the kind of data that can be monitored.
    \item Analysing these metrics may be misleading because these convey possible problems (\textit{i.e.}, the lack of resources) but do not point towards the actual problem (\textit{i.e.}, the faulty code).
\end{itemize}

\subsection{Example}
In a blog post, Stragliotto~\cite{Stragliotto-2019} provides an example of how THRON adopted this pattern. The author considers that \textit{"monitoring is the most important starting point to improve your product"}~\cite{Stragliotto-2019}, so they decided to update their monitoring architecture to improve their troubleshooting and alerting capabilities. In addition, they also improve their ability to analyse long-term trends and build dashboards.

The team decided to use Prometheus to collect infrastructure metrics to achieve their goal. 
In addition, they adopted Grafana \textit{"to query, visualise and generate alerts from our metrics"}~\cite{Stragliotto-2019}. It is very common in grey literature to find these two tools combined to set up a \textbf{Metrics Service}.

The author goes on to talk about the specific metrics that they chose to collect from their services. The post mentions the Four Golden Signals~\cite{Ewaschuk-SRE-2016} as the most important to monitor a system, but for the infrastructure, they followed the USE method~\cite{Gregg-2013}. Stragliotto defends that this method is better suited when the need is to \textit{"keep the physical resources under control"}~\cite{Stragliotto-2019}. Therefore, they collected utilization, saturation, and errors from many resources. Specific resources and metrics we could capture in this context include CPU utilization per core, memory usage, I/O wait time, and disk saturation. For example, using Prometheus' node exporter provides \textit{node\_cpu\_seconds\_total} and \textit{node\_disk\_io\_time\_seconds\_total} which can be used identify bottlenecks. Using Grafana, they built dashboards with thresholds and alerts for each metric, including stacked time-series plots and anomaly detection triggers
Note that, since infrastructure metrics convey the state of the infrastructure, they point to possible problems, but they are not the actual problem. This is a consequence that we consider very important for the pattern.

To conclude, Stragliotto declares that they \textit{"are happy about how [their] new architecture turned out, it works and it's starting to really help [them] keep [their] software under control"}. This view of the system was essential to make their detection and troubleshooting processes faster and fix their services more quickly.

\subsection{Known uses}
Stragliotto's~\cite{Stragliotto-2019} example described in the previous section is already a known use of this pattern that used Prometheus and Grafana to collect and visualise the utilisation, saturation and errors of their system's infrastructure.

An infrastructure software engineer post on callstats.io's blog~\cite{Callstats-2018} reports how the company uses Prometheus to monitor its services. They chose Prometheus for its seamless integration with Kubernetes, powerful query language and operational simplicity. The engineer states that the company's infrastructure and operations teams \textit{"use [Prometheus] to monitor resource usage and service performance, as well as data pipeline processing queues "}~\cite{Callstats-2018}.

Finally, a post by Campuzano on GumGum's Tech Blog~\cite{Campuzano-2019} explains how the company adopted Prometheus to monitor its systems. In addition, they utilised Grafana to visualise the data with customised dashboards. Although the author does not write about the actual metrics they collected, he mentions the usage of Prometheus' node exporter. This metrics exporter exposes \textit{"hardware and OS metrics on *nix systems"}~\cite{Campuzano-2019} and publishes more than 600 metrics. The author recognises that this amount of metrics can be a problem due to the high processing overhead they introduce.

\subsection{Related patterns}
The infrastructure metrics represent the state of the infrastructure. Analysing these logs helps notice symptoms of a problem. However, they are not usually enough to find what exactly is causing the issue. In such cases, the team should consider adopting \textsc{Application Metrics} to complement this pattern and get a full view of the system.

\section{Conclusions}\label{sec:conclusions}

Observability remains an important aspect of cloud-native architectures, enabling teams to diagnose failures, optimize performance, and maintain system reliability. The authors describe three design patterns to monitor cloud-native applications. \textsc{Distributed Tracing} supports end-to-end visibility across microservices by tracking requests as they propagate through the system, helping teams identify latency bottlenecks and failure points. \textsc{Application Metrics} suggests the collection and analysis of domain-specific performance indicators, allowing developers to proactively detect anomalies and optimize application behavior, and \textsc{Infrastructure Metrics} supports monitoring of underlying resources such as CPU, memory, and network usage, ensuring that cloud-native applications run efficiently and scale appropriately. 

These patterns are part of a collection of eleven design patterns presented previously by the authors~\cite{Albuquerque-Thesis-2022} and explained in Section~\ref{sec:about-patterns}. The patterns outlined in this paper are grounded in practice, but it would be interesting to empirically study how broadly they are understood and used in the way described in this paper. This can be a direction for future work, possibly leading to the refinement of the pattern descriptions, and assessment of their impact on real-world cloud-native systems.

\begin{credits}
\subsubsection{\ackname} We would like to thank Uwe Zdun, who helped us improve this paper significantly through the shepherding process, as well as all the participants in our writers' workshop at EuroPLoP 2025, who also provided a lot of thoughtful feedback---Daniel Reis, Diogo Maia, Francesco Urdih, Julia Pampus, Tiago Boldt Sousa, and Uwe Zdun.

This work is co-financed by Component 5 - Capitalization and Business Innovation, integrated in the Resilience Dimension of the Recovery and Resilience Plan within the scope of the Recovery and Resilience Mechanism (MRR) of the European Union (EU), framed in the Next Generation EU, for the period 2021 - 2026, within project HfPT, with reference 41.

\subsubsection{\discintname}
The authors have no competing interests to declare that are relevant to the content of this article.
\end{credits}

\bibliographystyle{splncs04}
\bibliography{references}
\end{document}